\documentclass[a4paper]{jpconf}
\usepackage[T1]{fontenc}
\usepackage{graphicx}

\renewcommand{\vec}[1]{\mathbf{#1}}
\begin{document}
\title{Microcavity quantum-dot systems for non-equilibrium Bose-Einstein condensation}

\author{I~M~Piper$^1$, P~R~Eastham$^2$, M~Ediger$^1$, A~M~Wilson$^1$, Y~Wu$^1$, M~Hugues$^3$, M~Hopkinson$^3$ and R~T~Phillips$^1$}
\address{$^1$ Cavendish Laboratory, University of Cambridge, J J Thomson Avenue, Cambridge CB3~0HE, UK}

\address{$^2$ School of Physics, Trinity College Dublin, Dublin 2, Ireland}

\address{$^3$ Department of Electronic and Electrical Engineering, University of Sheffield, Mappin Street, Sheffield, S1~3JD UK}

\ead{imp24@cam.ac.uk}

\begin{abstract}
We review the practical conditions required to achieve a
non-equilibrium BEC driven by quantum dynamics in a system
comprising a microcavity field mode and a distribution of localised
two-level systems driven to a step-like population inversion
profile. A candidate system based on eight 3.8nm layers of
In$_{0.23}$Ga$_{0.77}$As in GaAs shows promising characteristics
with regard to the total dipole strength which can be coupled to the
field mode.
\end{abstract}

\section{Introduction}
Experimental observation of polariton Bose-Einstein condensates
(BEC) has been reported in several different systems \cite{Kasprzak,
Balili, Bloch}. In all of these cases, the ground state is populated
by relaxation from an excited state, rather than populating the
ground state resonantly. A different approach to creating BEC in a
polariton system proposed by Eastham and Phillips \cite{Eastham},
does not involve dissipation or inelastic scattering. Instead the
condensation occurs as a result of the coherent dynamics of a
specific coupled photon-exciton system.

In general the proposed dynamical condensation could be implemented
in any system of discrete atomic-like states which are
dipole-coupled to a long-lived mode of the electromagnetic field. 
We consider the practical aspects of implementing this proposal in
semiconductor systems.

\section{Theoretical requirements}
\label{sec:theoreticalreq} The proposed experiment involves two
stages, which are separated in time and can be regarded as
independent. The first stage is the creation of a tailored exciton
population in the inhomogeneously-broadened exciton line using a
controlled pump pulse. For certain population profiles there is then
a second stage, in which the photon-mediated interactions between
the localized excitons lead to condensation. More specifically, the
normal modes of a population of dipole-coupled localized excitons
obey an equation [(\ref{eq:normalmodes})], which is essentially the
Cooper equation of BCS theory. It predicts that steps in the
energy-dependent exciton population result in new collective modes,
leading to condensation. The theoretical requirements for the
condensation stage are described by the equation ($\hbar=1$)
\begin{equation}
\label{eq:normalmodes}
  \lambda_\vec{k}-\omega_\vec{k}+ \kappa^2 n\int
  \frac{\nu(E)[2n(E)-1]}{\lambda_\vec{k}-E} dE=0.
\end{equation}
This equation determines the complex normal-mode frequencies
$\lambda_{\vec{k}}$ of the electromagnetic field, in a planar
microcavity containing an ensemble of localized dipole oscillators.
These normal modes are approximately plane waves, with in-plane
wavevector $\mathbf{k}$.  $\omega_\vec{k}$ is the frequency of the
bare cavity mode, whose imaginary part $-\gamma$ describes the
cavity losses. The localized exciton states are assumed to be
two-level systems, which can be either unoccupied or occupied.
$\nu(E)$ is the normalized distribution of the energies of these
transitions, and $n$ is their area density. Their average occupation
at energy $E$ is $n(E)$. 
Adiabatic rapid passage allows the creation of tailored population
profiles $n(E)$, with an energy resolution set by the duration of
the pump pulse~\cite{Eastham}.

The coupling $\kappa$ is the matrix element for a transition from
the unoccupied to the occupied state, assumed to be a constant for
simplicity. In the dipole gauge its explicit form is
\begin{equation}
  \kappa=\langle e r \rangle
  \sqrt{\frac{\hbar\omega}{2\epsilon_0\epsilon
      L_{\mathrm{eff}}}},
\end{equation}
where $\langle e r \rangle$ is the dipole moment of the transition,
$\hbar\omega$ its energy, and $L_{\mathrm{eff}}$ the effective width
of the cavity. A typical value for a semiconductor in the bandgap
region is $\langle e r \rangle\approx 20\;\mathrm{D}$, corresponding
to an electron moving approximately $0.5\;\mathrm{nm}$ 
\cite{Asada}. For $\hbar\omega\approx 1\;\mathrm{eV}$ and
$L_{\mathrm{eff}}\approx 20\;\mu\mathrm{m}$ this gives an estimate
$\kappa \approx 8\times10^{-7}\;\mathrm{meV\;cm^{-1}}$.

A solution to (\ref{eq:normalmodes}) with
$\lambda_\vec{k}^{\prime\prime}=\Im(\lambda_\vec{k})>0$ corresponds
to an exponentially growing normal mode, which leads to a large
population. Thus the presence of such solutions implies a dynamical
phase transition from an exciton population to a non-equilibrium
condensate. The character of this condensate depends on that of the
unstable normal mode. If it is essentially a pure electromagnetic
field mode we have a photon condensate, i.e. a laser. If it has a
significant excitonic polarisation component then we have
condensation of coupled light-matter states.

Close to a condensation threshold the imaginary part of
(\ref{eq:normalmodes}) is a gain-loss equation for the growth rate
of the mode:
$\lambda_\vec{k}^{\prime\prime}=-\gamma+\pi\kappa^2n\nu(\lambda_{\vec{k}}^{\prime})[2n(E)-1]$.
The energy range of the optically-active localized states is
approximately the inhomogeneous linewidth $\gamma_{\mathrm{inh}}$,
so typically $\nu(E)\sim 1/\gamma_{\mathrm{inh}}$. Thus a necessary
condition for condensation is, up to numerical factors of order one,
$\kappa^2n> \gamma\gamma_{\mathrm{inh}}$. A second requirement comes
from the real part of (\ref{eq:normalmodes}), which determines the
shifting of the normal mode energies due to the exciton population.
Large shifts indicate a significant exciton component, and hence
distinguish exciton-photon condensation from lasing. For realistic
$n(E)$ and $\nu(E)$ equation (\ref{eq:normalmodes}) gives shifts of
order $\kappa^2n/\gamma_{\mathrm{inh}}$. 
Thus if we require energy
shifts of order $E_{\mathrm{res}}$, as well as the gain-loss
criterion to be satisfied, we have $\kappa^2n> \gamma_{inh}
\max(\gamma,E_{\mathrm{res}})$.

An energy shift greater than the inverse of the exciton's
decoherence time implies an excitonic component to the condensate.
We hope to achieve somewhat larger shifts, of order
$1\;\mathrm{meV}$. This is about one order of magnitude smaller than
the vacuum Rabi splittings now routinely observed, and is a
reasonable benchmark for a new exciton-photon coupling phenomenon.
Assuming an inhomogeneous linewidth also of order $1\;\mathrm{meV}$,
and taking the estimate of the coupling strength above, gives an
order-of-magnitude estimate of the total area density required near
the field antinodes as $n > 10^{12} \mathrm{cm}^{-2}$.
The previous demonstration
of lasing in very similar vertical cavity surface-emitting
structures \cite{Saito} indicates that the required physical
r\'egime should be accessible.

\section{Candidate semiconductor systems}
\begin{figure}
\begin{minipage}[b]{0.45\linewidth}
\centering
\includegraphics[width=0.9\linewidth]{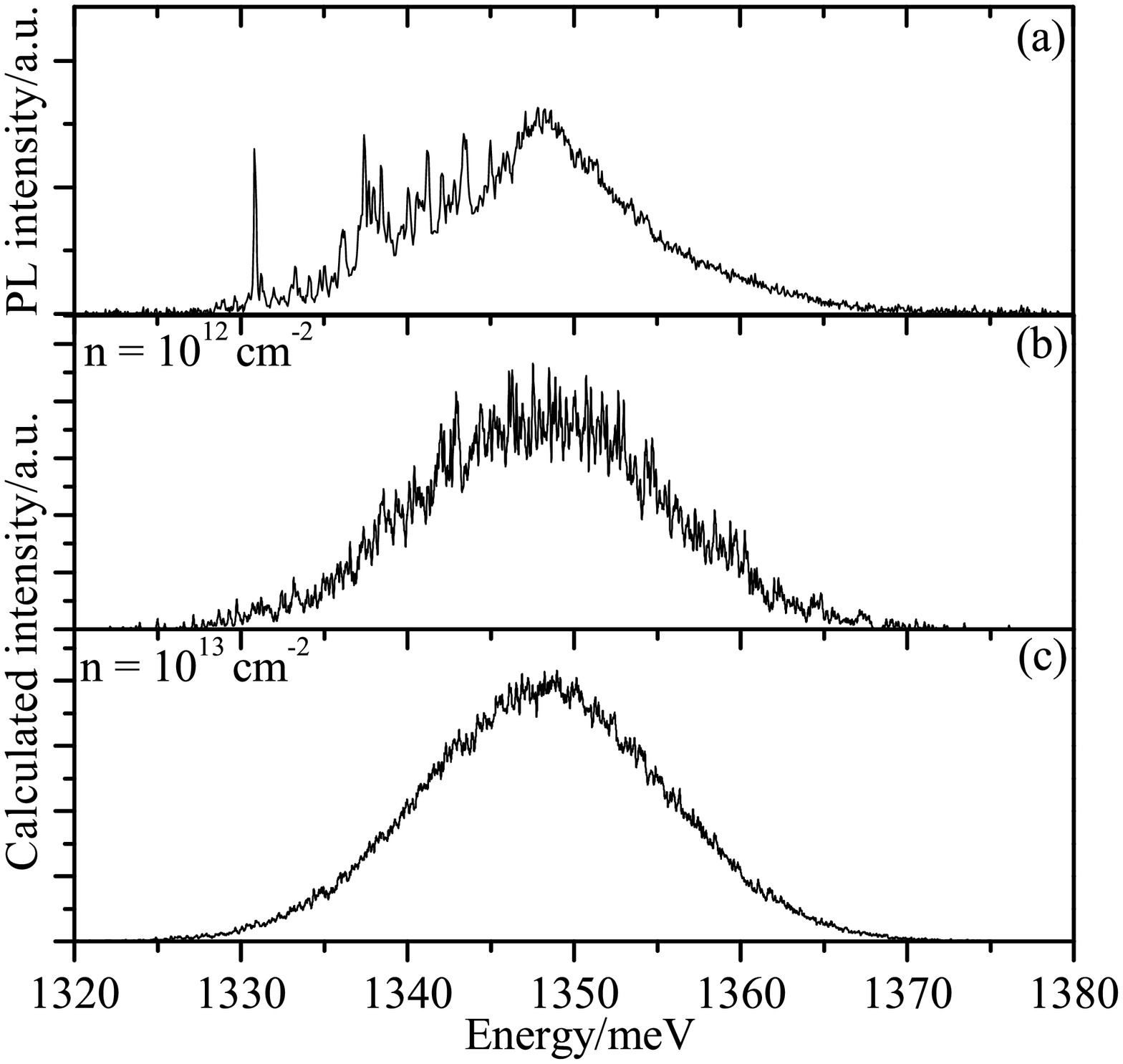}
\caption{(a) Confocal $\mu$PL from a region of area $\approx 1 \mu m^2$, showing both resolved and unresolved emission from the dot ensemble. Excitation was at 633nm, at a power density $\approx 10 \mathrm{W cm}^{-2}$. A model density of states of $n=10^{13}~\mathrm{cm}^{-2}$ (c) is too high to resolve individual transitions. Localised states can be resolved for an order of magnitude lower density ($10^{12}~\mathrm{cm}^{-2}$,~(b)), indicating that the density of states in the sample lies between that of (b) and (c).}
\label{Fig1}
\end{minipage}
\hspace{0.1cm}
\begin{minipage}[b]{0.45\linewidth}
\centering
\includegraphics[width=65mm]{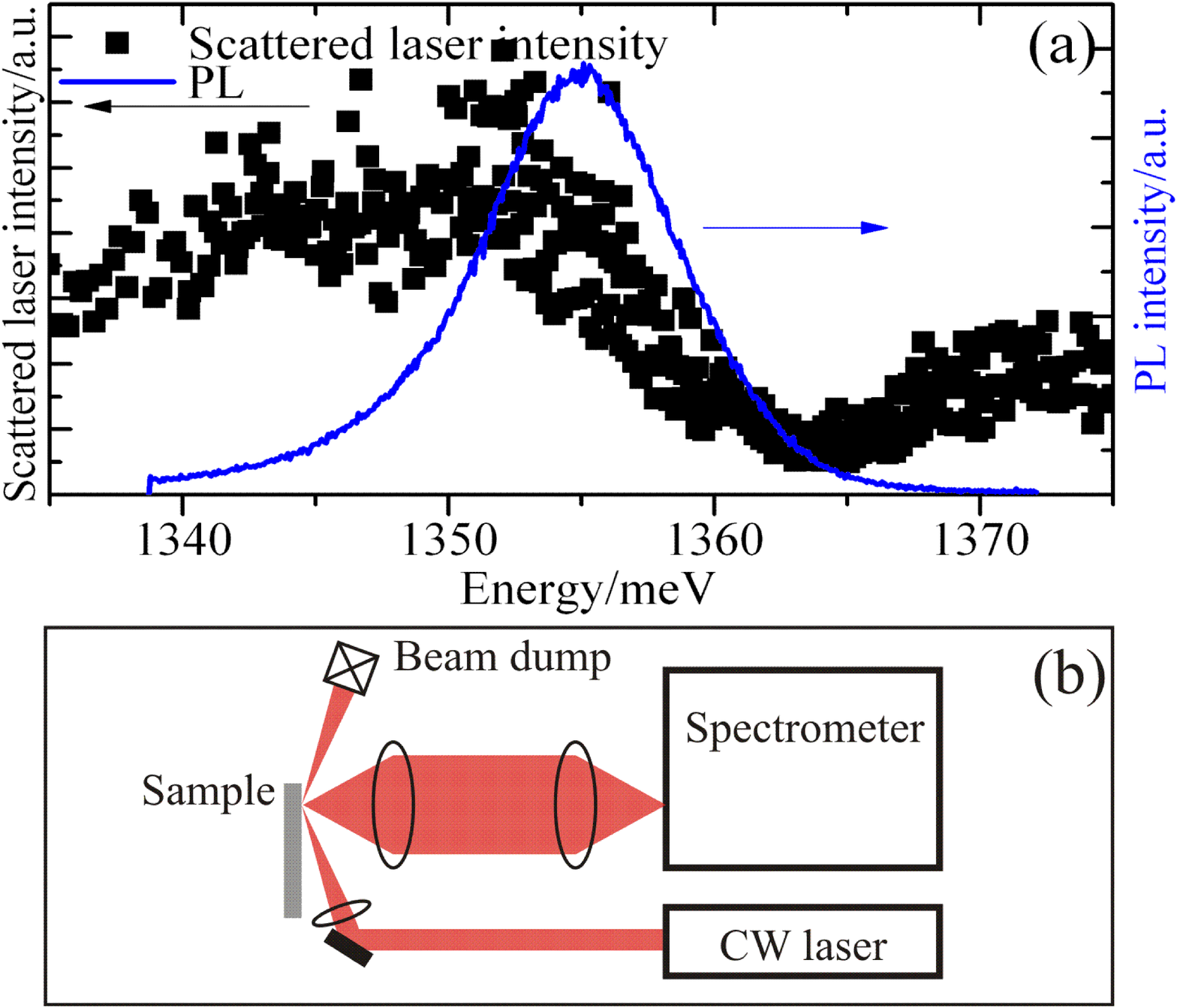}
\vspace{19.6pt}
\caption{(a) The spectral dependence of the weak RRS enhancement shows a dip on the high energy side of the line associated with reabsorption in the eight-layer stack. The low degree of RRS enhancemnent indicates 0D rather than 2D states. \\(b) The sample is excited at an oblique angle and the RRS normal to the sample collected. The scattered intensity as the CW laser is tuned across the transition is recorded.}
\vspace{2.88pt}
\label{Fig2}
\end{minipage}
\end{figure}

Stranski-Krastanow (S-K) dots are limited in density by the
requirements of the growth mode, and usually are strongly
inhomogeneously broadened, which reduces the spectral density. We
have therefore considered a system based on a "poor" quantum well,
in which alloy and thickness fluctuations localise the nominally 2D
states into quantum dots (QD). The key issues here relate to the
dipole coupling strength, density of dots, and proof of their
behaviour as isolated two-level systems. The sample consists of
eight 3.8nm layers of In$_{0.23}$Ga$_{0.77}$As with GaAs barriers,
grown by molecular beam epitaxy on an insulating GaAs substrate.
Microphotoluminescence ($\mu$PL) collects PL from an area $< 1 \mu
m^2$ and resolves emission from individual localised states, as
shown in Figure~\ref{Fig1}. By modelling the PL as a superposition
of Lorentzians in a Gaussian distribution we estimate the total
density of states (in eight layers) to be at least
10$^{12}~\mathrm{cm}^{-2}$ (Figure~\ref{Fig1}(b,c)). This system
therefore appears capable of satisfying the density requirement
implied in Section~\ref{sec:theoreticalreq}. A further important
feature of the PL evidence is that the total inhomogeneous linewidth
of 10meV is very well suited to the proposed cavity BEC experiment,
and is narrower than typically found for high density ($n
>10^{10}\mathrm{cm}^{-2}$) S-K dot samples.

Resonant Rayleigh scattering (RRS) is sensitive to localisation, and
for 2D states shows enhancement by a factor up to 100 on
resonance~\cite{Hegarty, Garro}. However, as localisation increases,
the intensity of the RRS signal decreases and becomes broader, as
the RRS efficiency depends on the ratio between the homogeneous and
inhomogeneous broadening of the transitions~\cite{Hegarty}. The weak
enhancement in the present samples (Figure~\ref{Fig2}(a)) is
consistent with a high degree of localisation.

\begin{figure}
\begin{minipage}[b]{0.45\linewidth}
\includegraphics[width=0.9\linewidth]{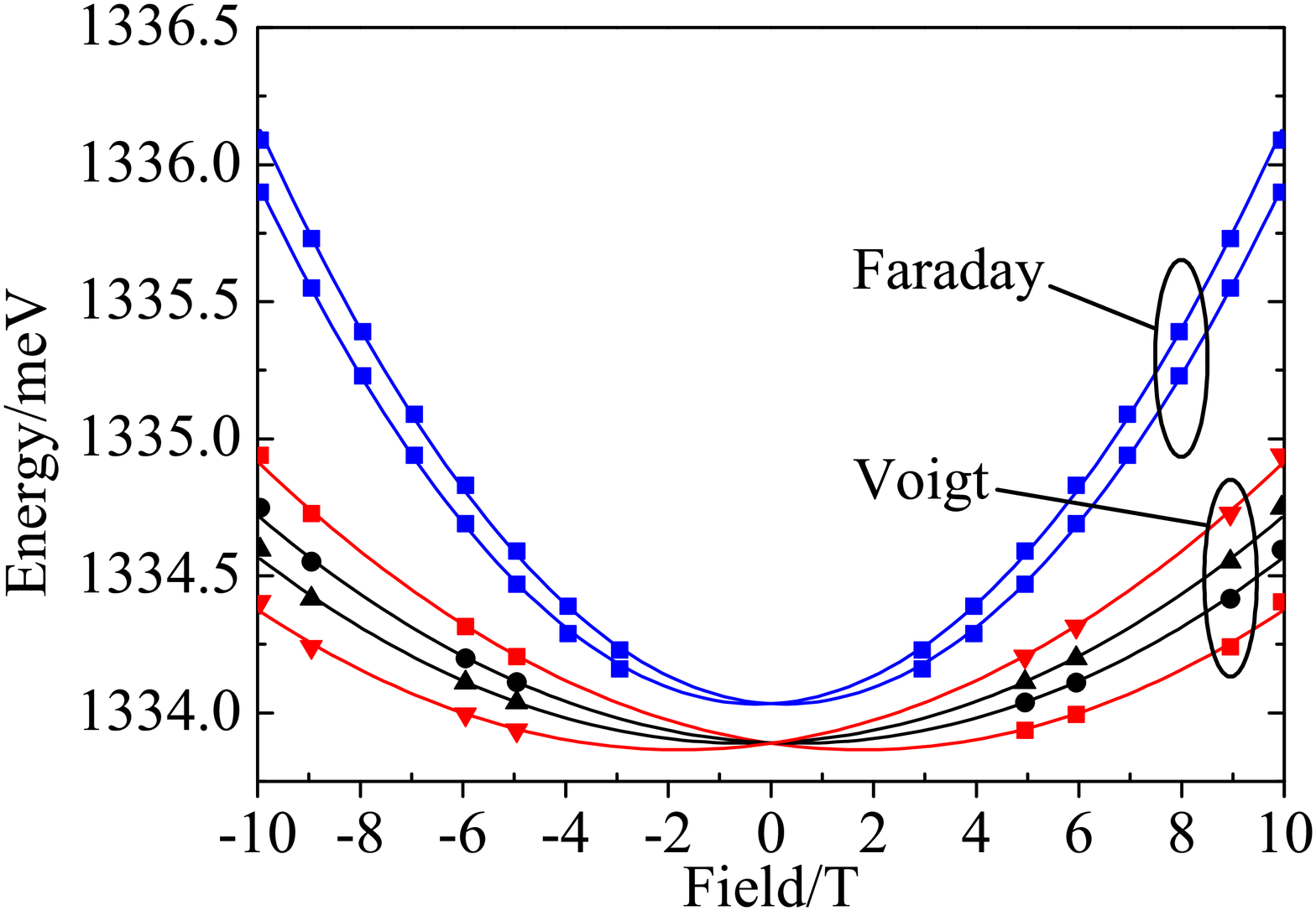}
\caption[font=small]{Diamagnetic shifts and spin fine structure of a single localised state in both the Faraday and Voigt geometries, illustrating the change in \emph{e-h} correlation length as the field is moved from parallel to perpendicular to the growth axis. Markers show every tenth data point; lines show a least-squares minimised regression fit.}
\label{Fig3}
\end{minipage}
\hspace{0.1cm}
\begin{minipage}[b]{0.45\linewidth}
\centering
\includegraphics[width=0.9\linewidth]{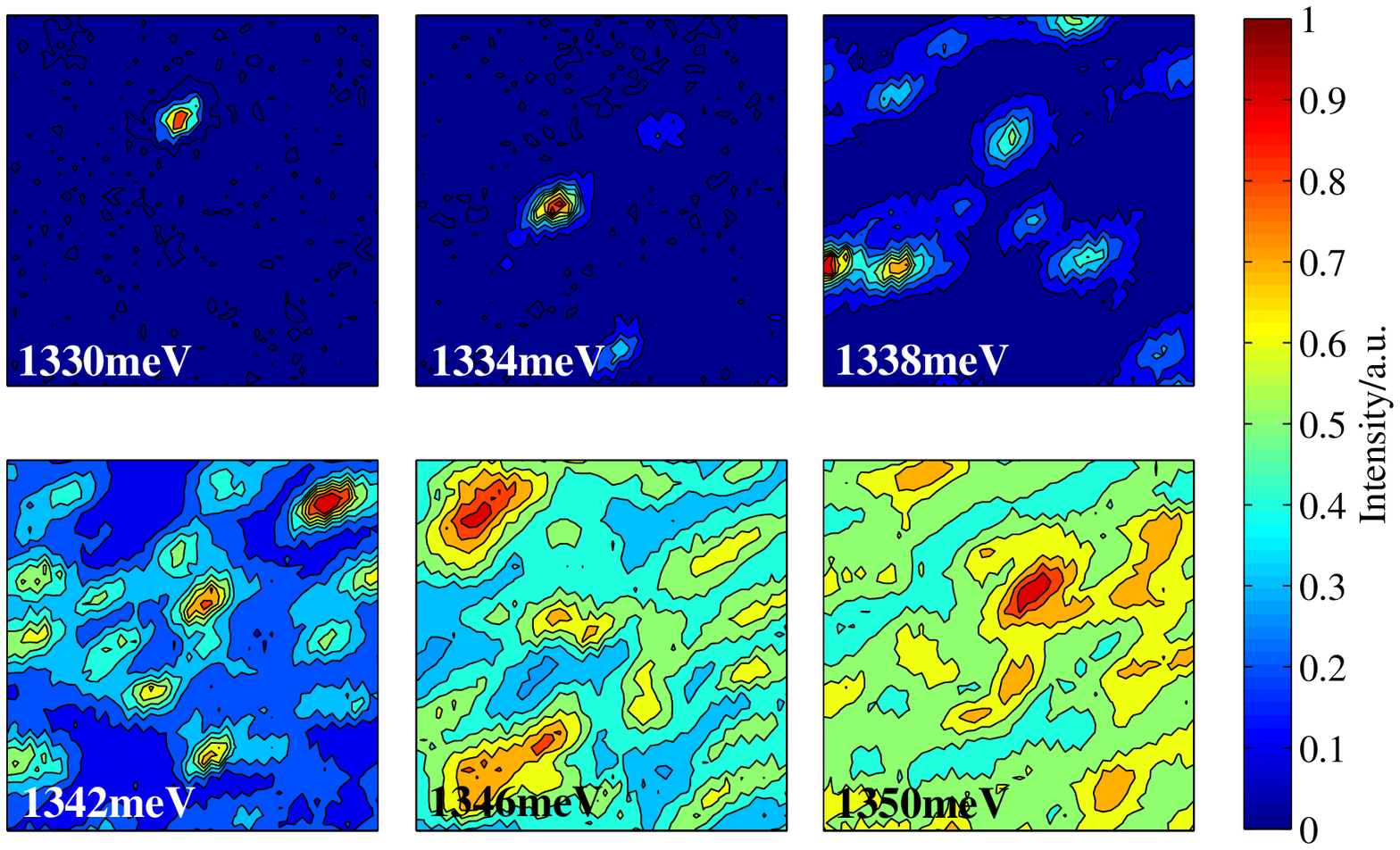}
\vspace{17pt}
\caption{Map of $\mu$PL emission from a $15 \mu m \times 15 \mu m$ region at six different energies. }
\vspace{64pt}
\label{Fig4}
\end{minipage}
\end{figure}

The electron-hole correlation length $\rho$ is given by $ \langle
\rho^2 \rangle = \frac{8 \gamma_{2} \mu}{e^2}$, where $\gamma_2 B^2$
is the diamagnetic shift and $\mu$ the reduced exciton
mass~\cite{Phillips}. The diamagnetic coefficient of one state
measured using magneto-optical $\mu$PL \cite{Kehoe} is found to be
20.1~$\mu$eV~T$^{-2}$ in the Faraday geometry and
7.6~$\mu$eV~T$^{-2}$ in Voigt, shown in Figure~\ref{Fig3}. We
estimate the extent of the \emph{e-h} correlation length to be
approximately 7nm and 4nm respectively. As this is a simple model,
these values are consistent with those of localised states in QDs.
Electron microscopy of a QD system with similar composition shows
the QDs are of length 40-100nm and width 20-30nm; these states have
similar diamagnetic shifts to those measured in the present
work~\cite{Mensing}.

$\mu$PL emission maps at six photon energies are shown in
Figure~\ref{Fig4} for a $15 \mu m \times 15 \mu m$ region indicating
elongation of the emitting regions in the $[1\bar{1}0]$ direction.
The combination of these experiments indicates that the localised
states in narrow In$_{0.23}$Ga$_{0.77}$As/GaAs quantum wells are
promising candidates for the proposed dynamically-driven BEC, as
they have a high density of discrete localised states distributed
over a suitably narrow energy range.

\section{Conclusions}
We have outlined the theoretical origin of the requirements for a
suitable system in which to realise the proposed condensation
scheme. Narrow In$_{0.23}$Ga$_{0.77}$As/GaAs quantum wells show a
suitably high density of localised states to offer a good system for
attempted observation of the dynamically-driven BEC.

\ack{This work was conducted with financial support from EPSRC grant EP/F040075/1 and the Science Foundation Ireland grant 09/SIRG/I1592. We thank the staff of the Cavendish Laboratory Workshops for their expert technical assistance.}


\section*{References}
\begin{thebibliography}{99}
\bibitem{Kasprzak}
Kasprzak J \emph{et al.} 2006 \emph{Nature} {\bf 443} 409

\bibitem{Balili}
Balili R, Hartwell V, Snoke D, Pfeiffer L, West K 2007
\emph{Science} {\bf 316} 1007

\bibitem{Bloch}
Bajoni D, Senellart P, Lema\^{i}tre A and Bloch J 2007 \emph{Phys.
Rev} B {\bf 76} 201305

\bibitem{Eastham}
Eastham P R and Phillips R T 2009 \emph{Physical Review} {\bf B79}
165303 

\bibitem{Asada}
Asada M and Suematsu Y, 1985, \emph{IEEE J. Quantum Electronics}
{\bf 21} 434

\bibitem{Saito}
Saito H, Nishi K, Ogura I, Sugo S and Sugimoto Y 1996 \emph{Appl.
Phys. Lett} {\bf 69} 3140

\bibitem{Hegarty}
Hegarty J, Sturge M D, Weisbuch C, Gossard A C and Wiegmann W 1982
\emph{Phys. Rev. Lett.} {\bf 49} 930

\bibitem{Garro}
Garro N \emph{et al.} 1997 \emph{Phys. Rev.} B {\bf 55} 13752

\bibitem{Phillips}
Phillips R T, Steffan A G, Newton S R, Reinecke T L and Kotlyar R
2003 \emph{phys. stat. sol.} (b) {\bf 238} 601

\bibitem{Kehoe}
Kehoe T, Ediger M, Phillips R T, and Hopkinson M 2010 \emph{Rev.
Sci. Instrum.} {\bf 81} 013906 

\bibitem{Mensing}
Mensing T, Reitzenstein S, L\"{o}ffler A, Reithmaier J P, and
Forchel A 2005 \emph{Physica} E {\bf 32} 131






\end{thebibliography}
\end{document}